\documentstyle[12pt,aasms4]{article}
\def \yskip{\penalty-50\vskip3pt plus 3pt minus 2pt}
\def \pp{\par \yskip \noindent \hangindent .4in \hangafter 1}
\def \abc#1#2#3#4 {\pp#1, {\sl#2}, {\bf#3}, #4}
\def \blank {\lower 5pt\hbox to 0.75in{\hrulefill}}
\def \kms{km~$\rm{s}^{-1}$}
\def \cc{$\rm{cm}^{-3}$}
\def \lam{$\lambda$}
\def \Ha{H$\alpha$}

\def \mum{$\mu$m}

\def \msol{M$_{\odot}$}

\newfont{\rten}{cmr10} 
\def\arcdeg{\hbox{$^\circ$}}
\def\arcmin{\hbox{$^\prime$}}
\def\arcsec{\hbox{$^{\prime\prime}$}}
 
\begin{document}
\normalsize
\title{ISOCAM Molecular Hydrogen Images of the Cep E Outflow\footnotemark}
 
\author{Alberto Noriega-Crespo}
\affil{Infrared Processing and Analysis Center,}
\affil{California Institute of Technology and Jet Propulsion Laboratory}
\affil{Pasadena, California, 91125}
\affil{Electronic mail: alberto@ipac.caltech.edu}

\author{Peter M. Garnavich}
\affil{Harvard-Smithsonian Center for Astrophysics, Cambridge, MA 02138}
\affil{Electronic mail: peterg@cfanewton.harvard.edu}

\affil{\&} 

\author{Sergio Molinari}
\affil{Infrared Processing and Analysis Center,}
\affil{California Institute of Technology and Jet Propulsion Laboratory}
\affil{Pasadena, California, 91125}
\affil{Electronic mail: molinari@ipac.caltech.edu}

\footnotetext{ISO is an ESA project with the instruments funded by ESA 
Members States (especially the PI countries: France, Germany, the Netherlands
and the United Kingdom) and with the participation of ISAS and NASA.}

\begin{abstract}
The physical characteristics of Cepheus E (Cep E) `embedded'
outflow are analyzed using ISOCAM images in the v=0-0 S(5) 
6.91 \mum~ and S(3) 9.66 \mum~ molecular hydrogen lines. We find that 
the morphology of the Cep E outflow in the ground vibrational H$_2$ lines 
is similar to that of the near infrared v=1-0 2.12 \mum~line. At these 
mid-IR wavelengths, we do not detect the second H$_2$ outflow which is almost 
perpendicular to Cep E 2.121 \mum~flow or traces of H$_2$ emission along 
the second $^{12}$CO $J = 2-1$ outflow at $\sim 52$\arcdeg~angle, down to a 
surface brightness of 12 - 46 $\mu$Jy/arcsec$^2$.

We do detect at 6.91 \mum~the likely source of the main H$_2$ and CO outflows, 
IRAS 23011+6126, and show that the source is easily seen in all IRAS bands 
using HiRes images. The source is not detected at 9.66 \mum, but we think 
this agrees with the interstellar extinction curve  which has a minimum at 
$\sim$ 7 \mum, but rises a $\sim 9.7$ \mum~due to the strong absorption 
silicate feature, enhanced in this case by a cocoon surrounding the 
Class 0 object. This idea is supported by our models of the spectral energy
distribution (SED) of the central object. The models assume that 
the main source of opacity is due to bare silicates and our best fit for
the SED yields a total mass of envelope of 17 \msol~and a dust temperature 
of 18 K.
\end{abstract}

\section{Introduction}

The Cepheus E (Cep E) outflow is an excellent object
to study the relationship between the physical properties of the optical
stellar jets and those of embedded outflows. The existence of a molecular CO
flow originating in that region was first indicated by 
\markcite{fuk89}Fukui (1989) in his catalog of molecular outflows, 
but it was the K$^\prime$ image of \markcite{hod94}Hodapp (1994), from 
his imaging survey of molecular outflows, that brought Cep E
to the forefront. The south lobe of Cep E is observed optically in
\Ha\ and [SII] \lam\lam6717/31 (\markcite{anc97}Noriega-Crespo 1997), 
and it has been named Herbig-Haro (HH) object 337 (\markcite{dev97}Devine, 
Reipurth \& Bally 1997). The spectra of this lobe displays one of the lowest 
ionizations measured in HH objects, with a ratio [SII]/\Ha = $8.8\pm 0.2$ 
(\markcite{aya98}Ayala et al. 1998). 
 
The Cep E is bright in H$_2$ emission lines at 2\mum~and is likely to be 
driven by the IRAS 23011+6126 source, presumably a class Class 0 object 
(\markcite{eis96}Eisl\"offel et al. 1996). Because of its complex morphology 
and strong H$_2$ emission, Cep E has been used to gauge the reliability of 
3-D molecular jet models (\markcite{sut97}Sutter et al. 1997). There is also
clear evidence for a second outflow in H$_2$ almost perpendicular to the 
main flow emission, as indicated by some faint H$_2$ knots 
(\markcite{eis96}Eisl\"offel et al. 1996). Perhaps more surprising is the 
presence of a second molecular CO outflow, detected in both $^{12}$CO $J=2-1$
and $^{12}$CO $J=1-0$ molecular lines (\markcite{lad97}Ladd \& Hodapp 1997;
\markcite{lad97b}Ladd \& Howe 1997), at an angle of $\sim 52$\arcdeg\ from 
the main CO flow, i.e. apparently unrelated to the second H$_2$ outflow.
The $^{12}$CO emission indicates terminal velocities of 80 \kms~and 
$-$125 \kms~in its north and south lobes respectively, which implies a 
dynamical age of  $\sim 3\times 10^3$ years, at a distance of 730 pc 
(\markcite{eis96}Eisl\"offel et al. 1996). The absence of a second source 
using interferometric observations at 2.65 mm suggests that both outflows 
arise very close to the position of IRAS 23011+6126 source 
(\markcite{lad97}Ladd \& Howe 1997). 

In this study we present new ISOCAM images of the Cep E outflow taken in the 
ground vibrational level (v=0-0) H$_2$ lines S(5)at 6.91 \mum~and S(3) 
at 9.66 \mum.  The motivation of this work was to use the ground vibrational 
H$_2$ lines, which are predicted to be much brighter than the v=1-0 S(1) 
line (\markcite{wol91}Wolfire \& K\"onigl 1991), to study the excitation
across the outflow (using the S(3)/S(5) ratio), and to try to detect the 
second H$_2$ outflow and the central source at mid-infrared wavelengths, as 
well as to search for traces of H$_2$ emission along the second CO outflow.
These observations are complemented by HiRes IRAS images.

\section{Observations}

The Cep E outflow has a projected size onto the plane of the sky of 
$\sim 1\arcmin$, so the ISOCAM images were obtained using a 2$\times$3 CVF 
raster map with a 6\arcsec~FOV pixel scale and 30\arcsec~steps. 
Four CVF filters were used; two very close centered on the v=0-0 S(3) 9.665 
\mum~and S(5) 6.909 \mum~H$_2$ lines respectively (see below); plus two 
nearby continuum CVF steps at 9.535 \mum~and 6.855 \mum.  
We selected the S(3) 9.665 \mum~line,
despite the fact that its wavelength is right in the middle of the strong 
silicate absorption feature at $\sim 9.7$ \mum, because the line is
expected to be strong and should help to constrain the depth of
the silicate absorption in the models of the spectral energy distribution
(SED).
 
The ISOCAM data was deglitched using the Multi-resolution Median 
Transform Method (CIA). The detectors transient effects were treated 
using the IPAC Model (Ganga 1997). The target dedicated time (TDT) was of
2728 seconds, with twelve stabilization time steps on line and ten in the 
continuum prior to the on target observations. The TDT was spent with 2/3 
of the time on line and 1/3 in the continuum, this means approximately
900 secs in each of the H$_2$ lines. The ADU fluxes were transformed 
into calibrated fluxes using the upgraded values of the system response
(e.~g. ISOCAM Observer's Manual, Tables 12-17). These values are given 
in ADU/sec/mJy/pixel and correspond to 125.52 (step 227 \@ 9.660 \mum), 
65.80 (step 331 \@ 9.535 \mum), 122.12 (step 21 \@ 6.911 \mum) and 121.686 
(step 22 \@ 6.855 \mum). The FWHM of four the filters are
(9.660, 9.535, 6.911, 6.855) \mum~ = (0.27, 0.22, 0.17, 0.17) \mum;
and we used an integration time step of 2 secs and a ADC gain of 2.

We present also for comparison a near infrared image at the v=1-0 S(1) 
2.12~\mum~obtained at the 3.5m Apache Observatory with a 256$\times$256 
array at f/5 with a 0.482\arcsec per pixel scale. A complete analysis of the
imaging and spectroscopic near infrared data is presented elsewhere 
(Ayala et al.  1998). HiRes IRAS images are presented also to support these 
observations.  These images have one degree field of view with 15\arcsec~pixels
and can reach a spatial resolution of $\sim 1\arcmin$. The HiRes images have 
been processed using Yu Cao algorithm (see e.~g. Noriega-Crespo et al. 1997).

\section{Results}

\subsection{Morphology}

The grayscale images of H$_2$ at 6.91 \mum~and 9.66 \mum~are
presented in Figures 1 and 2 respectively. From these images is evident that
the morphology of these different molecular hydrogen lines is very similar. It
is also clear that the main difference between them is the central intensity 
peak at  6.96 \mum. This intensity peak coincides with the position of the 
IRAS 23011+6126 source, as determined using interferometric observations by 
\markcite{eis96}{Eisl\"offel et al. (1996), i.~e.
$\alpha$(2000) = 23h 03m 13.0s, $\delta$(2000) = 61\arcdeg~42\arcmin~
26.5\arcsec. The IRAS 23011+6126 source is also detected in the continuum 
frame at 6.855 \mum, indicating that the emission is not dominated by the 
excited H$_2$ emission. The Figure 3 shows the 6.91\mum~image with an overlay 
of the H$_2$ 2.12 \mum~emission, and once again the morphology of both 
molecular lines is very similar. 

One way to understand these observations is in terms of the interstellar
extinction law (see e.~g.\markcite{mat90} Mathis 1991), which it has a minimum 
at $\sim 7$ \mum~and the rises to a peak at $\sim 10$ \mum, where this maximum
is due to the silicate absorption feature. If the IRAS 23011+6126 is embedded 
in a dusty
envelope, as expected for a Class 0 source, then the absorption by silicates 
will be even larger and enough to swamp the H$_2$ emission at 9.66 \mum. 
The IRAS 23011+6126 source appears in all the IRAS bands, as is shown in 
Figure 4, which displays IRAS HiRes maps at 12, 25, 60 and 100 \mum~centered 
on the Cep E source.

We recall that another two outflows have been detected around the Cep E
source. One outflow is observed in the 2.12 \mum~H$_2$ emission and is
almost perpendicular to the main H$_2$ flow. The second one is detected in the
$^{12}$CO $J = 2-1$ transition and is centered in the IRAS 23011+6126 
source ($\pm$2.3\arcsec), with an orientation of $\sim 52$\arcdeg~with respect
the main H$_2$ flow and a scale of $\sim 4$\arcmin~(\markcite{lad97}Ladd \& 
Hodapp 1997). We don't detect in the H$_2$ lines at 9.66 \mum~or 6.91 \mum~any 
signatures of the second H$_2$ outflow nor of the second CO outflow observed 
at millimeter wavelengths. For the faint 2.12 \mum~H$_2$ outflow this perhaps 
is not surprising since the S(3) 6.91 \mum~line would have needed to be 
$\sim 70$ times stronger than S(1) 2.12 \mum~line (based on simple
C-type shock models (e.g.\markcite{smth95}Smith 1995)) to overcome its very 
low surface brightness in comparison with the brightest regions.
We estimate that the RMS noise levels of our ISOCAM images, 
based on measurements of the background are approximately 
46 $\mu$Jy/arcsec$^2$ at 9.66 \mum, and 12 $\mu$Jy/arcsec$^2$ at 6.91 \mum.  
The values of the minimum contours in Figures 1b and 2b (continuum subtracted)
are 0.4 and 0.2 mJy/arcsec$^2$ for the 9.66 \mum~and 6.91 \mum~lines 
respectively, so a factor 70 times lower in brightness for the faint 
H$_2$ outflow is at the noise level.
There are not traces in the IRAS HiRes images of the faint H$_2$ flow nor of
the CO outflow. The HiRes images do not resolve either the second source near
IRAS 23011+6126, as is shown in Figure 4. The 12 and 25 \mum~images in Figure 4
display only point sources, and although there is some faint emission at 
60 \mum~and 100 \mum~ at a PA$\sim -45$\arcdeg, this is probably be due to 
the diffuse emission from the nearby sources. The existence of a second
source at 2$\arcmin$ SW from the IRAS 23011+6126 source has recently been 
confirmed by \markcite{tst98}L. Testi (1998) with OVRO observations at 1.3 mm.
 
\subsection{Fluxes}

 As mentioned before the $v = 0 - 0$ S(3) and S(5) H$_2$ were selected for 
our observations because plane parallel molecular shock models predict,
for the conditions encountered in HH objects, that these lines could be 
$\sim 10 - 100$ stronger than the $v = 1 - 0$ S(1) 2.12 \mum~line if they
are collisionally excited. Molecular shock models specifically calculated for 
HH objects (\markcite{wol91}Wolfire \& K\"onigl 1991) considered as typical 
parameters e.~g. a shock velocity of $v_s = $25 \kms, an initial preshock gas 
density
of $n_0 = 10^3$ \cc~and a preshock magnetic field of $B_0 = 30$ $\mu$G, a 
molecular hydrogen abundance ${n_{H_2}\over n}\sim 0.5$, an atomic hydrogen 
abundance ${n_{H}\over n}\sim 3\times 10^{-3}$, a neutral gas temperature of
$T\sim 2000$ K and an electron temperature $T_e \sim 3000$ K.
The point of ennumerating some of these parameters (there are a few 
more) is to illustrate the difficulty in comparing shock models with 
observations and to stress that the models should be taken as a guide,
since the physical conditions across a shock front change and are
more complex.

For the input parameters of the shock models mentioned
above, the expectation is that the ratios of the H$_2$ lines should be
${0-0~S(3)\over 1-0~S(1)} \sim 157$ and ${0-0~S(5)\over 1-0~S(1)} \sim 28$, 
or essentially ${0-0~S(3)\over 0-0~S(5)} = 5.6$. Figure 5 shows the
ratio of our S(5) to S(3) H$_2$ images, which indicates the the ratio
across the outflow is nearly unity. The ratio is constant except for 
the region around the IRAS 23011+6126 source, which is not detected in 
the S(5) 9.66 \mum~image, and at the edge of the outflow, which is probably 
due to an artifact produced by the  mismatch between the elliptical Gaussian 
function use to smear out the S(3) 6.91 \mum~ image, and the true shape of
the first Airy ring produced by the ISOCAM optics.

Nearly constant distributions across the outflow have been also measured for 
the ${2-1~S(1)} \over {1-0~S(1)}$ and ${3-2~S(3)}\over {1-0~S(1)}$ ratios
\markcite{eis96}(Eisl\"offel et al. 1996). This behavior of the ratios is very
difficult to explain with simple shock models or shock geometries, and in the
case of the v=2-1 S(1) and v=3-2 S(3) lines ratios, the best results were 
obtained with C-type bow shocks. Such C-type bow shock model needed shock
 velocities of $\sim 200$ \kms~ and preshock densities of $\sim 10^6$ \cc~
(\markcite{eis96}Eisl\"offel et al. 1996), which are larger than what is 
measured in most optical outflows.

A constant v=0-0 S(5) to S(3) ratio also indicates the lack of a steep
extinction gradient between the north and south lobes, and that the extinction
is mostly important around the IRAS 23011+6126 source. This is interesting
because we know that the south lobe is visible at optical wavelengths
(\markcite{anc97}Noriega-Crespo 1997), while the north lobe is not.

Finally, we have measured the flux of the IRAS 23011+6126 source at 
6.855 \mum~, set an upper limit for the 9.535 \mum~flux (which may be still 
affected by the silicate feature) and measured the IRAS fluxes from the HiRes 
images. In Figure 6 we show the SED
of the source, including the values obtained at 1.25 mm
and 2.22 \mum~by \markcite{lef96}Lefloch et al. (1996) and 2.65 mm by 
\markcite{lad97b}Ladd \& Howe (1997). We have overplotted (following 
\markcite{lad97b}Ladd \& Howe 1997) four simple gray body models, at 
T$_{dust}$ = 10, 20, 30 and 40 K, of the form 
$F_\nu = B_\nu(T)~(1 - e^{-\tau})~\Omega_s$ with $\tau \propto \nu^{-2}$ 
and the optical depth is normalized to the 2.65 mm flux.
The simple models do at reasonable job at the IRAS and millimeter wavelengths
for  T$_{dust}$ = 20 - 30 K, but are unable to fit the shorter wavelengths.

It is possible to build more sophisticated models for the SED
which include a density and temperature structure for the 
cloud core or envelope which surrounds the Class 0 object
(\markcite{and94}Andr\'e \&  Montmerle 1994), and which also take into account
 the dust opacity as a function of chemical composition and wavelength 
(see Appendix). 
In Figure 7 we display three models with different inner gas cloud core
densities ($6\times 10^4$, 8$\times 10^4$ and $10^5$ \cc), but otherwise 
identical in their input parameters. The model with an initial 
density of 8$\times 10^4$ \cc~(solid line) fits the observations amazingly 
well and is consistent with 
the upper limits at 2.22 \mum~and 9.855 \mum, which are based 
in `no detections'. The models also illustrate how well they fit
the IRAS and sub/millimeter observations, and how sensitive they are to
the presence of the $\sim 20$ \mum~silicate feature. This feature essentially 
disappears in the $6\times 10^4$ \cc~model (dotted line) and not surprisingly 
becomes deeper with an increasing density at $10^5$ \cc~(dashed line).

The model SEDs in Figure 7 assume a dust opacity dominated by bare silicates, 
hence the strong absorption features at 10 and 20 \mum, 
and a dust temperature of 18 K. The models consider a power law
density and temperature distributions (see Appendix), with a core inner
radius of 0.065 AU, and outer radius of 0.1 pc. The total masses for the cloud
envelope surrounding the IRAS source for the three models (increasing in
density) are 13, 17 and 22 \msol. Our best model (with T$_{dust}$ = 18 K
and M$_{env}$ = 17 \msol) yields different values than those obtained by 
using a constant density  distribution and the simple gray body models, i.~e.
T$_{dust}$ = 20 K and M$_{env}$ = 10 \msol~(\markcite{ladb}Ladd \& Howe 1997). 

\section{Conclusions}

 We have analyzed some of the physical characteristics of Cep E embedded
outflow using ISOCAM images in the v=0-0 S(5) 6.91 \mum~and S(3) 9.66 
\mum~molecular hydrogen lines. We find that the morphology of Cep E outflow
of the ground vibrational H$_2$ lines is similar to that of the near infrared 
emission in the v=1-0 2.12 \mum~line. At these wavelengths, and at surface
brightness of 12 - 46 $\mu$Jy/arcsec$^2$, we do not detect the second 
H$_2$ outflow almost perpendicular to the main 2.12 \mum~flow, nor we found 
traces of H$_2$ along the second $^{12}$CO $J = 2-1$ outflow at 
$\sim 52$\arcdeg~angle.

We detect at 6.91 \mum~the likely source of the main H$_2$ and CO outflows, 
IRAS 23011+6126, and show that the source is well detected in all IRAS bands 
using HiRes images. The source is not detected at 9.66 \mum (nor at 9.54 \mum),
but we think this 
agrees with the interstellar extinction curve which has a minimum at 
$\sim 7$ \mum, but rises a $\sim 9.7$ \mum~due to the strong silicate 
absorption feature, further enhanced in this case by a cocoon surrounding the 
Class 0 object as the model of the SED seems to indicate.

The ${0-0~S(5)}\over {0-0~S(3)}$ ratio is uniform and near unity across the 
outflow, a fact which is difficult to explain with simple plane parallel shock
models (\markcite{eis96}Eisl\"offel et al 1996). A constant S(5) to S(3) ratio
also indicates that the extinction, with the exception around the IRAS 
23011+6126 source, is not affecting the H$_2$ emission of the 
outflow lobes at these wavelengths.

Assuming that the main source of opacity around IRAS 23011+6126 is due to bare 
silicates, then our best model of SED for the envelope surrounding 
the Class 0 source yields a total mass of 17 \msol~and a dust temperature 
of 18 K, 

\acknowledgements
We thank Ken Ganga for his help and insight on the ISOCAM calibration and
the data reduction procedures. Our gratitude goes also to Jochen Eisl\"offel 
for sharing with us the analysis of Cep E with ISO. We thank N. King for 
obtaining the imaging data, and the referee for helpful comments and a careful
reading of the manuscript.

\clearpage

\begin{center}
{\bf Appendix}
\end{center}

The envelope around Cep E source is modeled as a series of spherically
symmetric dust shells, where temperature and density are assumed to
vary according to radial power laws whose exponents are free
parameters. Once the external radius of the envelope is fixed, the
inner envelope boundary is also determined and is equal to the radius
where dust attains its sublimation temperature ($\sim$1500 K).  As our
aim was not only to provide a reasonable model for the global spectral
energy distribution of Cep E source, but also to verify if the non detection
at 9.67 \mum~ could be due to silicate absorption, we then decided to
adopt the dust opacities tabulated by \markcite{oh94}Ossenkopf \& Henning 
(1994) for an MRN (\markcite{mrn77} Mathis, Rumpl \& Nordsieck 1977) silicate 
dust with variable volumes of ice mantles, instead of the usual assumption 
(e.g. \markcite{h83} Hildebrand 1983) of a dust opacity simply expressed as 
a power law of frequency. 

Dust emission is computed from each shell as:

\begin{equation}
F_{\nu, i}= \kappa_{\nu} B(\nu, T_i) \rho_i V_i
\label{flux}
\end{equation}

where $\kappa_{\nu}$ is the dust mass opacity, $B(\nu, T)$ is the Planck
function, $\rho_i$ is the density and V$_i$ is the volume of $i^{th}$
shell. First, for each frequency the optical depth is computed inward
as $\tau=\sum _i (\kappa_{\nu} \rho_i dr_i)$ until either $\tau=1$ is
eventually reached for a certain shell, or the inner envelope radius is
reached; having identified the inner shell which contributes to
observed radiation, the emitted flux is computed outward according to
Eq.~\ref{flux} until the external radius is reached. The flux from each
shell is extincted by the intervening shells toward the observer, and
finally the contributions from all shells are summed at each
frequency.

\clearpage

\begin{center}
Figure Captions
\end{center}
 
\figcaption[noriega_crespo_fig01.ps]{
(a) Grayscale image of the Cep E outflow in the v=0-0 S(3) 9.66 \mum~
H$_2$ line (with continuum) in a 2\arcmin~field of view.
(b) The same grayscale image (continuum subtracted) with contours beginning at 
0.4 mJy/arcsec$^2$, and increasing by a factor $2^{1/2}$.
\label{f1}}

\figcaption[noriega_crespo_fig02.ps]{
(a) Grayscale image of the Cep E outflow in the v=0-0 S(5) 6.91 \mum~
H$_2$ line (with continuum) in a 2\arcmin~field of view.
(b) The same grayscale image (continuum subtracted) with contours beginning at 
0.2 mJy/arcsec$^2$, and increasing by a factor $2^{1/2}$.
\label{f2}}

\figcaption[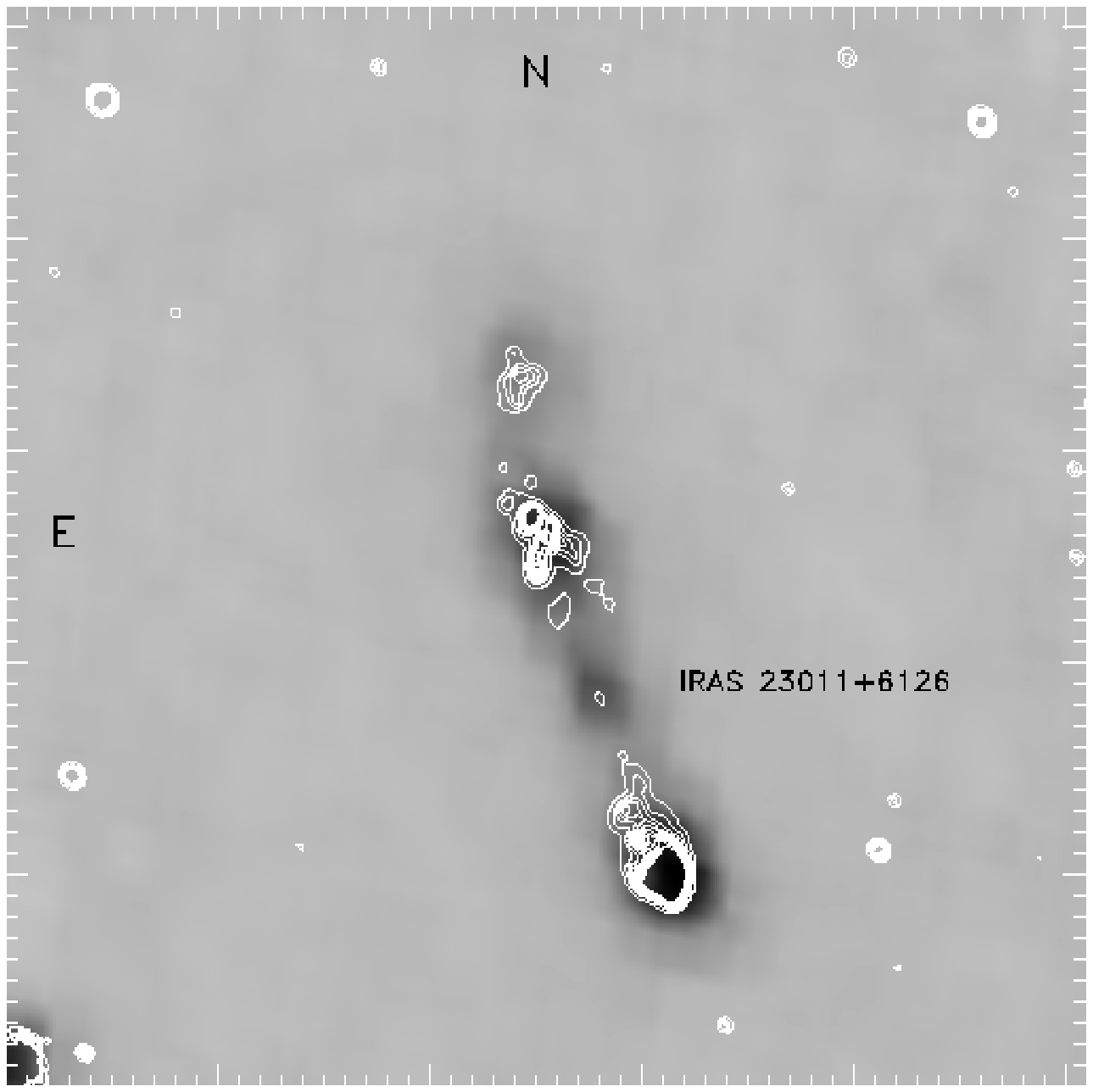]{
Grayscale image of the v=0-0 S(5) 6.91 \mum~line (with continuum)
with a linear contour map of the v=1-0 2.12~\mum~emission.
\label{f3}}

\figcaption[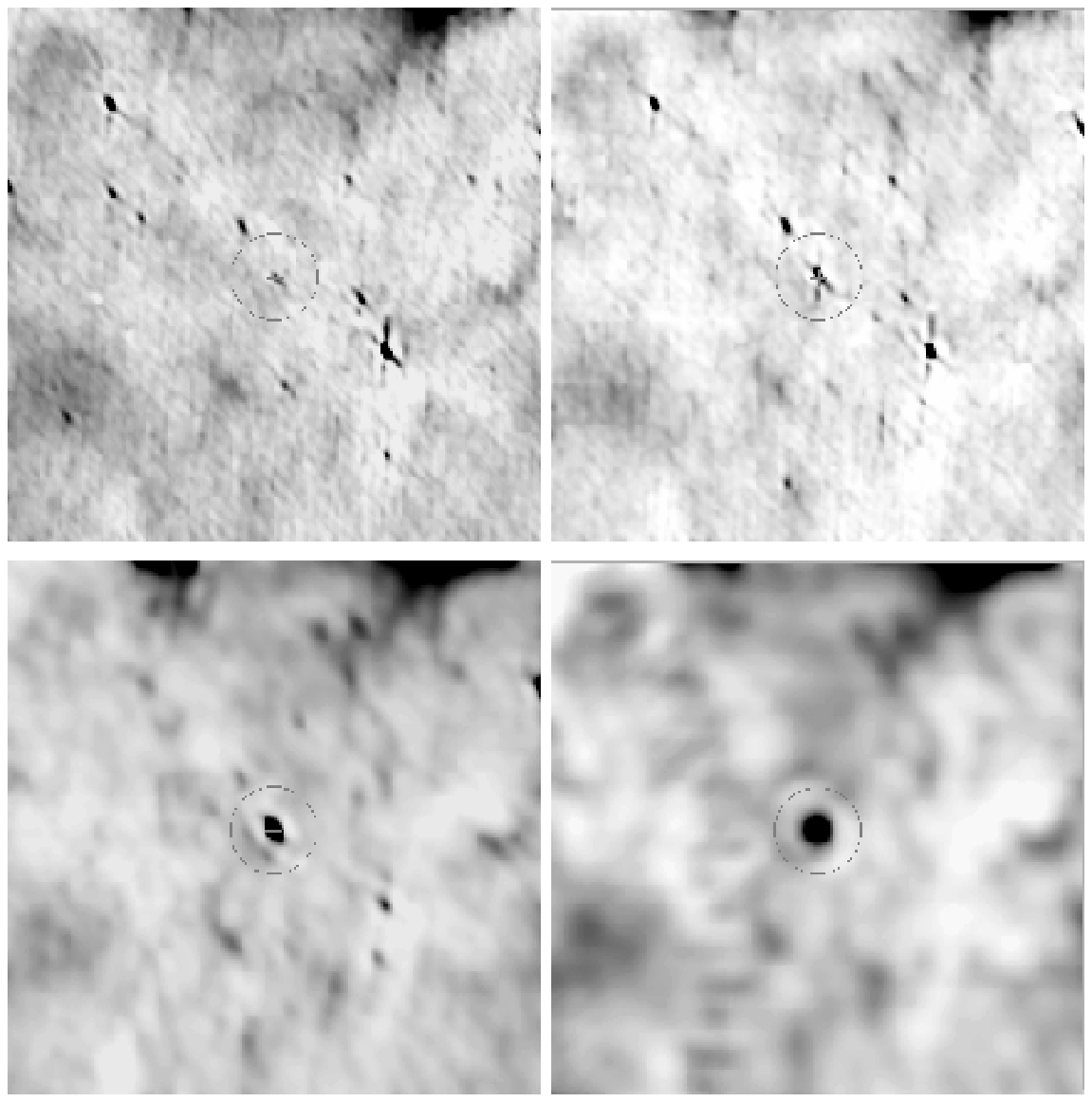]{
IRAS HiRes images at (from top left to right) 12, 25, 60 and 100 \mum
centered on the IRAS 23011+6126 source.
The field is one degree, with the usual orientation, and the circle
surrounding the IRAS source has a 5 \arcmin~ radius.
\label{f4}}

\figcaption[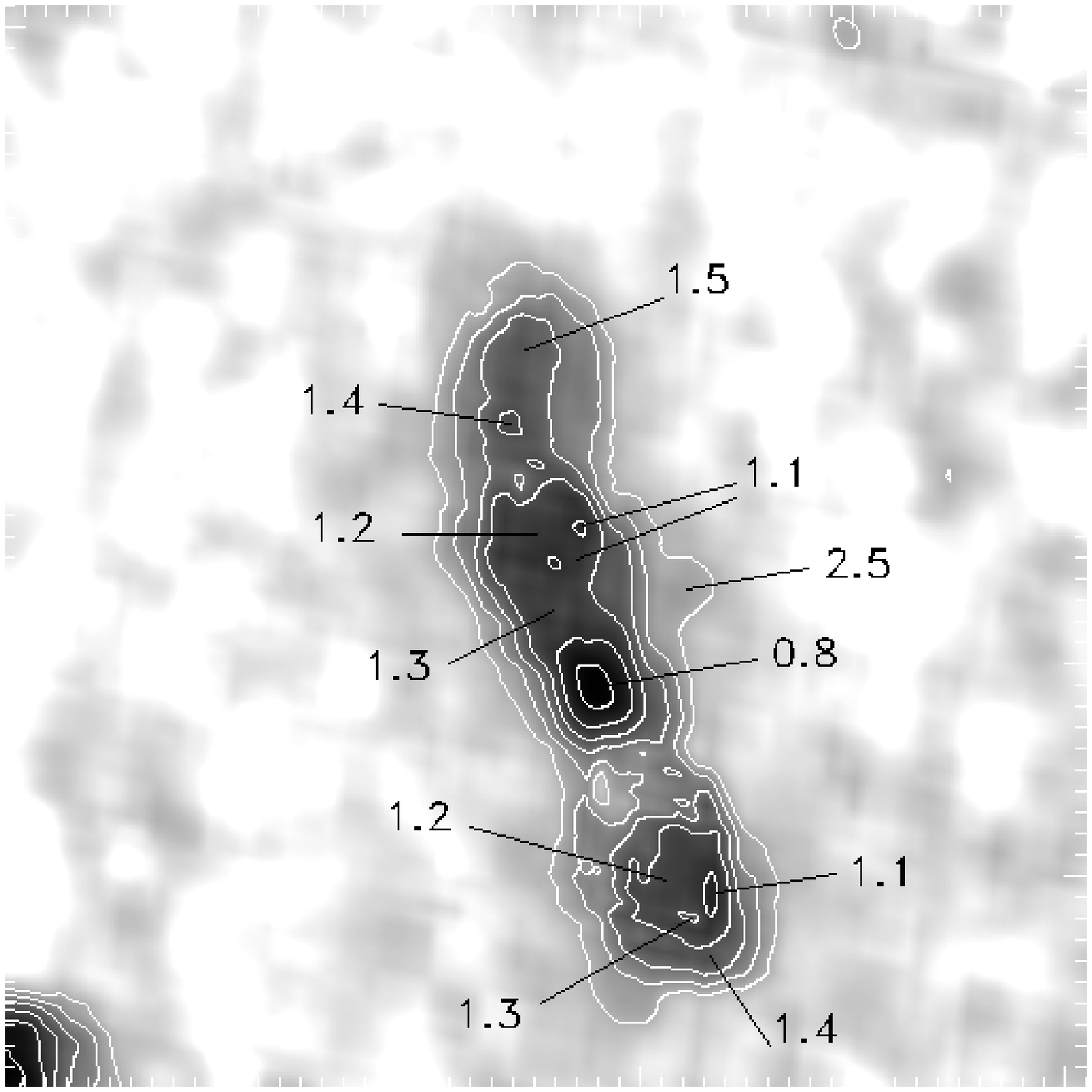]{
The v=0-0 S(3) 9.66 \mum~ to v=0-0 S(5) 6.91 \mum~ratio across the
Cep E outflow. The ratio is essentially constant and near unity, except
around the IRAS 23011+6126 source which is no detected at 9.66 \mum, and
at the edges due to an artifact.
\label{f5}}

\figcaption[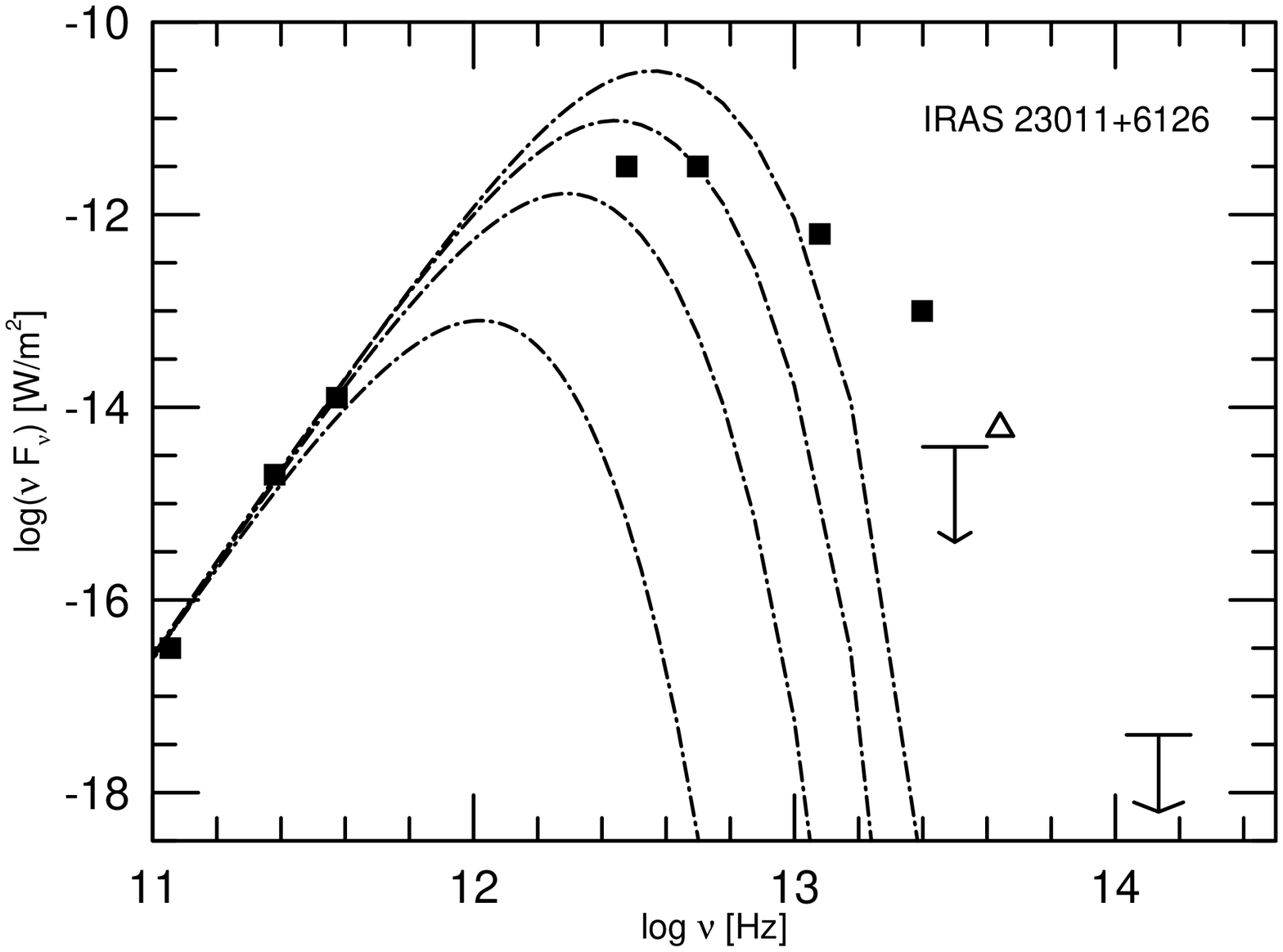]{
The spectral energy distribution of IRAS 23011+6126 source (following
Ladd \& Howe 1997). The triangle corresponds to our 6.855 \mum~measurement,
and the upper limits are for the 9.535 \mum~and 2.22 \mum~integrated fluxes.
The IRAS fluxes were measured from the HiRes images. 
Overplotted for comparison four simple gray body models with T$_{dust}$ = 10, 
20, 30 and 40 K, normalized at the 2.65 mm flux
\label{f6}}

\figcaption[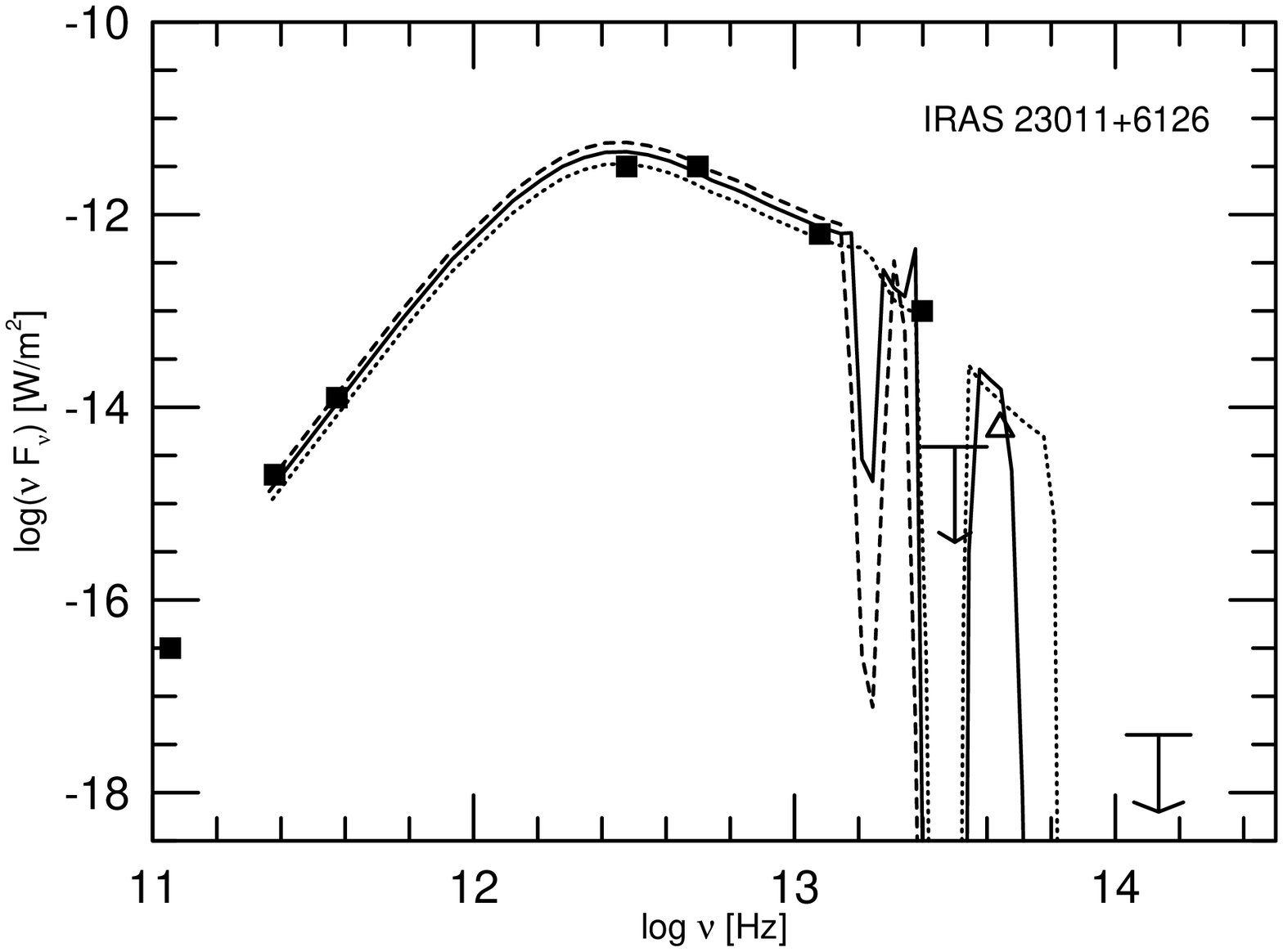]{
More detailed models of the SED for IRAS 23011+6126 source using the same data
as in Figure 6. The models assume a dust opacity  dominated 
by dust grains made of bare silicates, hence the strong 
absorption features at 10 and 20 \mum~(see the appendix for details).
The input parameters of the models are identical except for the initial
inner core gas density of $6\times 10^4$ \cc~(dotted line), $8\times 10^4$ \cc~
(solid line) and $10^5$ \cc~(dashed line).
\label{f7}}

\clearpage

\begin{figure}[v]
\plotone{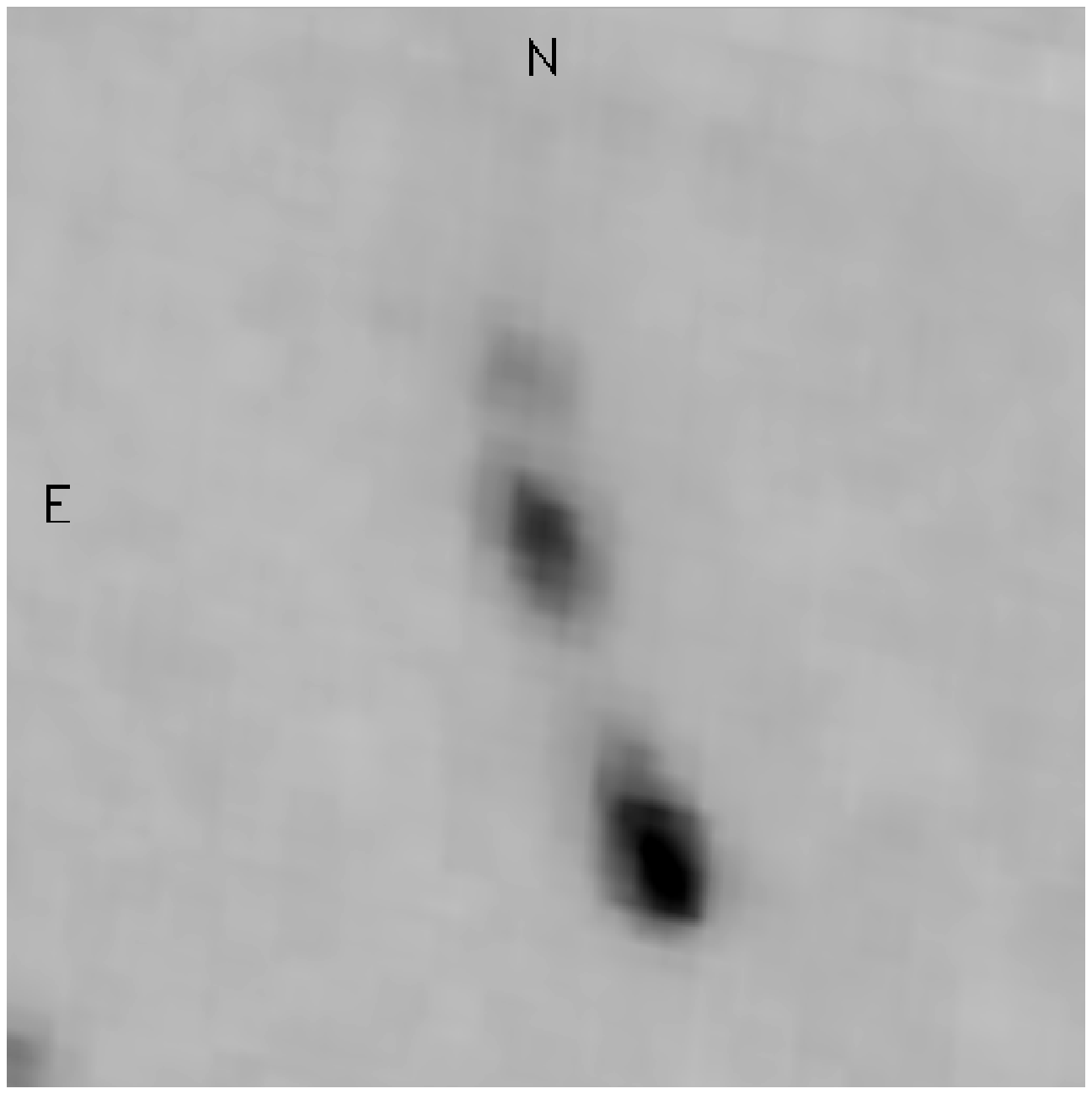}
\end{figure}

\clearpage

\begin{figure}[v]
\plotone{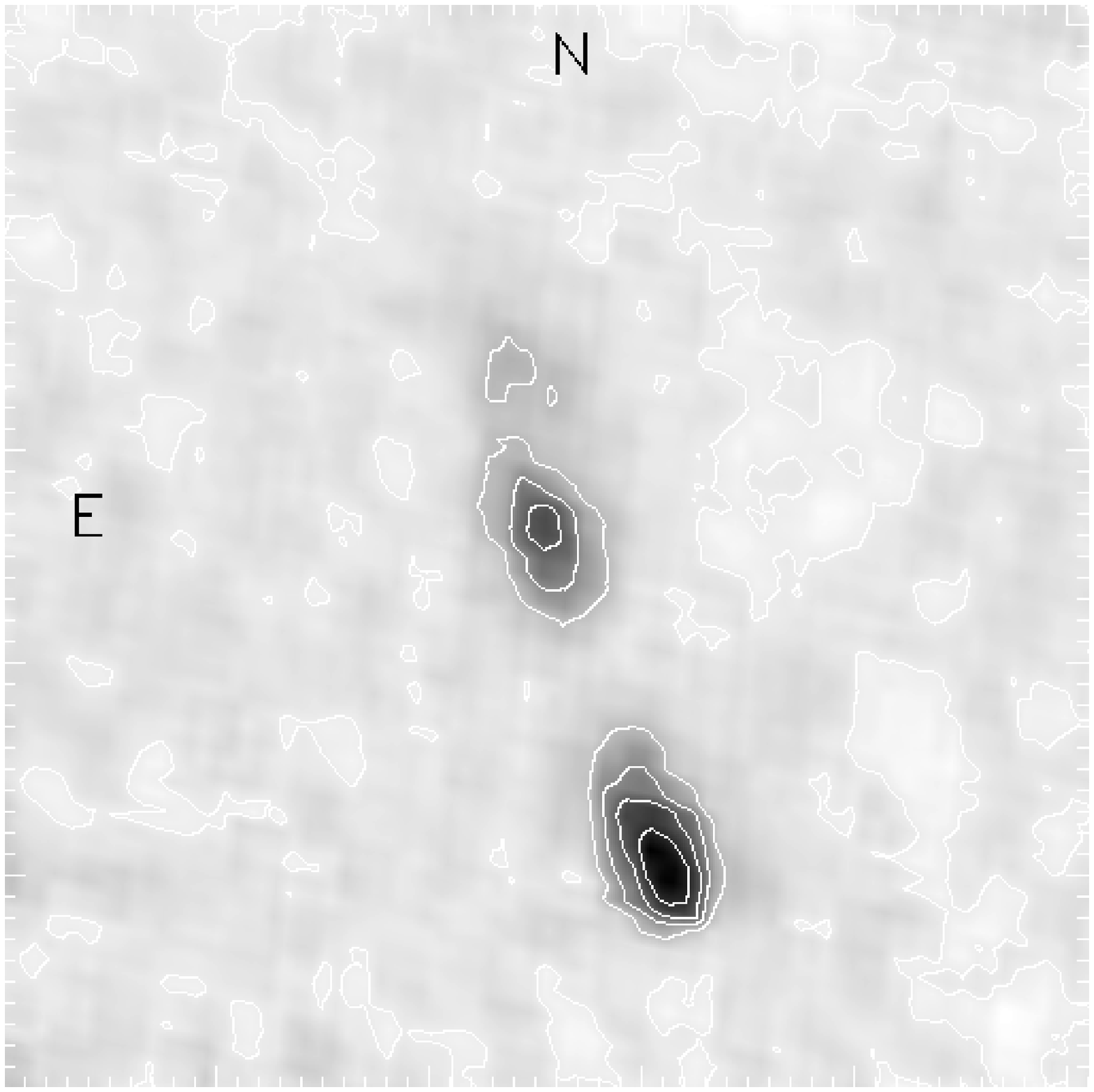}
\end{figure}

\clearpage

\begin{figure}[v]
\plotone{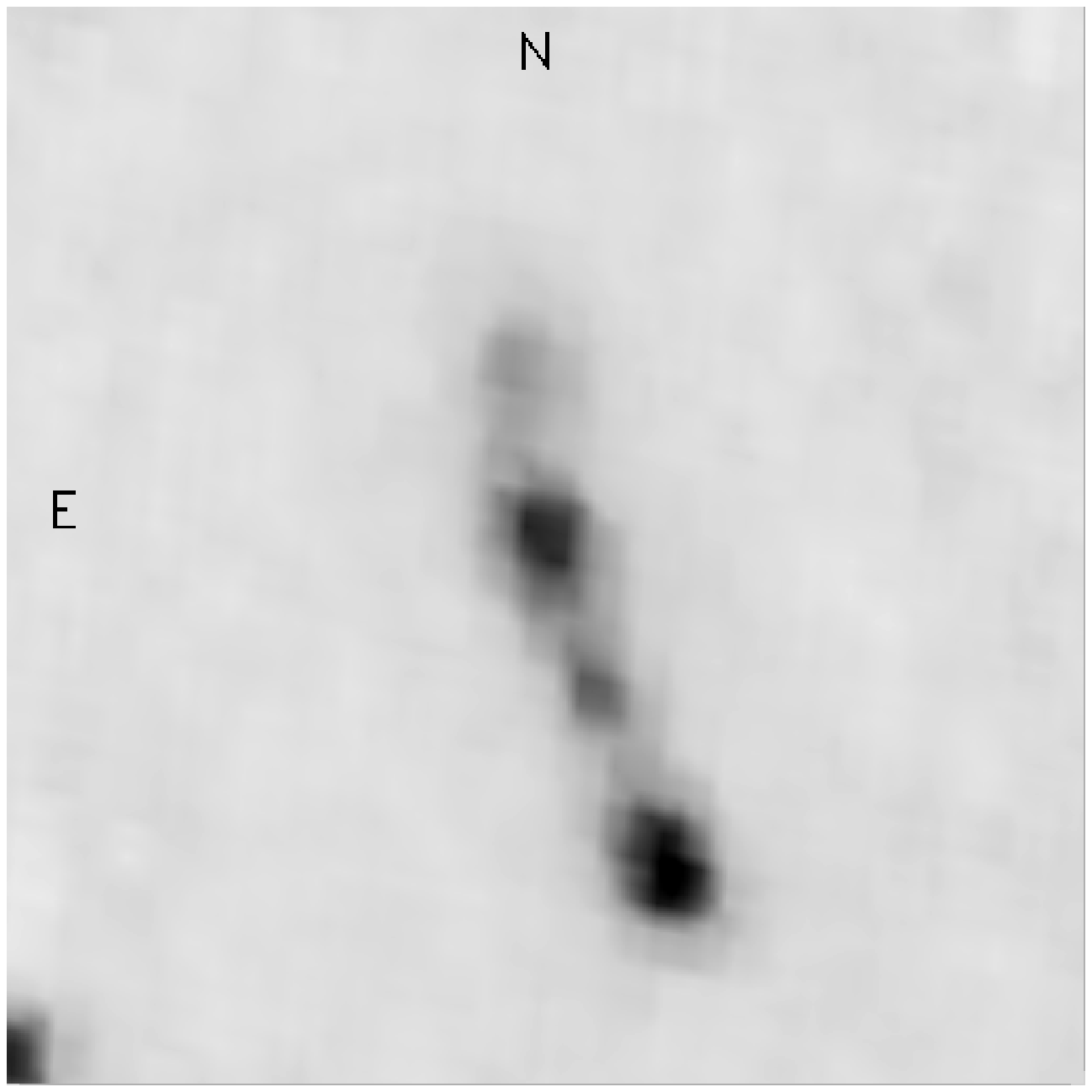}
\end{figure}

\clearpage

\begin{figure}[v]
\plotone{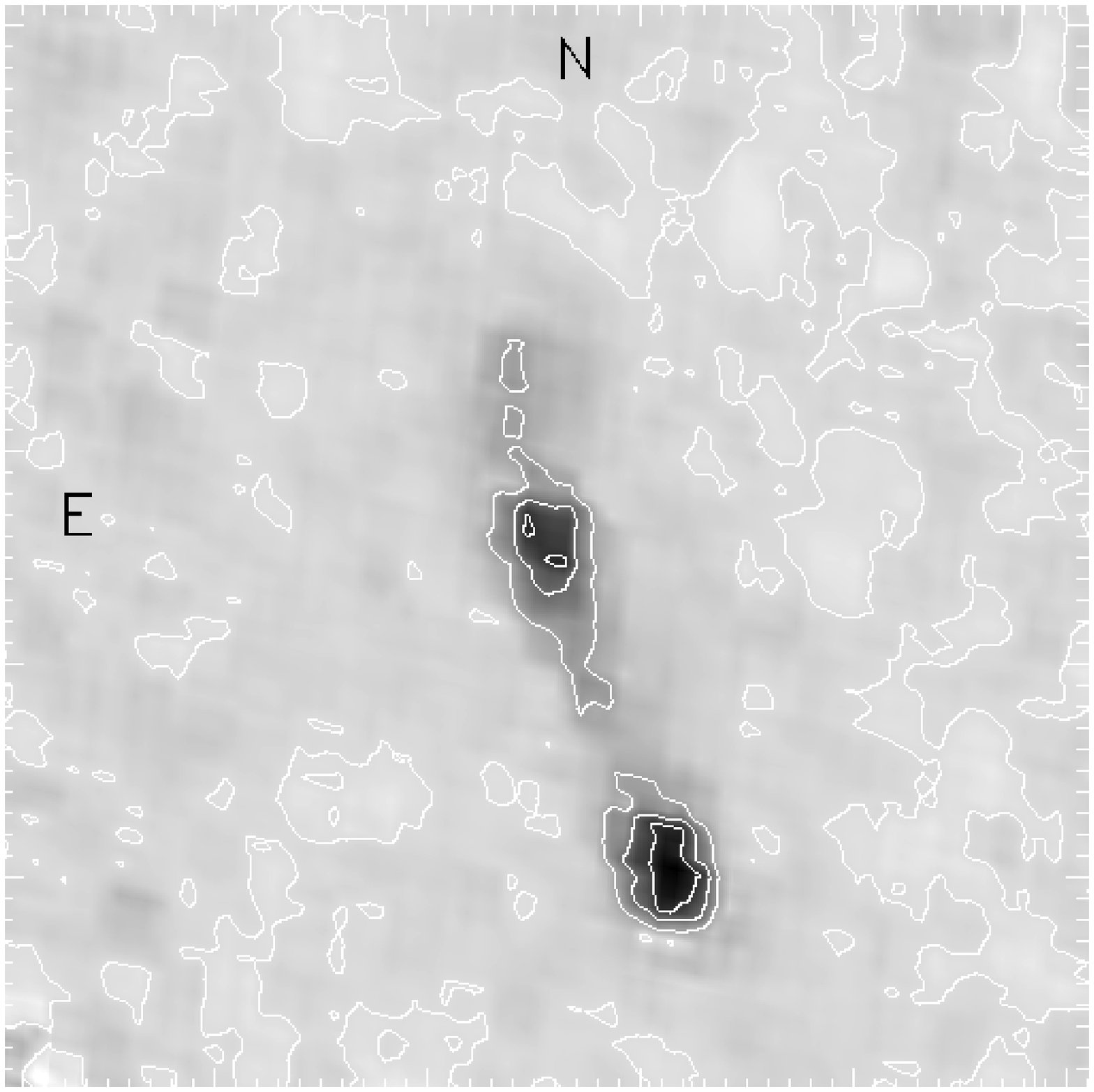}
\end{figure}

\clearpage

\begin{figure}[v]
\plotone{noriega_crespo_fig03.ps}
\end{figure}

\clearpage

\begin{figure}[v]
\plotone{noriega_crespo_fig04.ps}
\end{figure}

\clearpage

\begin{figure}[v]
\plotone{noriega_crespo_fig05.ps}
\end{figure}

\clearpage

\begin{figure}[v]
\plotone{noriega_crespo_fig06.ps}
\end{figure}

\clearpage

\begin{figure}[v]
\plotone{noriega_crespo_fig07.ps}
\end{figure}

\clearpage

\begin{references}

\reference{and94}Andr\'e, P., \& Montmerle, T. 1994, ApJ, 420, 837

\reference{aya98}
Ayala, S., Garnavich, P., Curiel, S., Noriega-Crespo,
A., Raga, A.C., \& B\"ohm K.H. 1998 (in preparation)

\reference{dev97}
Devine, D., Reipurth B., \& Bally J. 1997 in
``Low Mass Star Formation from Infall to Outflow'' Ed. F. Malbet \& 
A. Castets, p91

\reference{eis96}
Eisl\"offel, J., Smith, M.D., Davis, C.J., \& Ray, T.P. 1996, AJ, 112, 2086

\reference{fuk89}
Fukui, Y. 1989, in ``Low Mass Star Formation and Pre-Main Sequence 
Objects'', ESO 1989, ed. B. Reipurth, p95

\reference{gan97} 
Ganga, K. 1997, Private Communication

\reference{hod94}
Hodapp, K.-W. 1994, ApJSS, 94, 615


\reference{h83}
Hildebrand, R.H. 1983, QJRAS, 24, 267


\reference{lad97}
Ladd, E.F., \& Hodapp, K. 1997, ApJ, 474, 749

\reference{lad97b}
Ladd, E.F., \& Howe, J.E. 1997 in ``Low Mass Star Formation from Infall to
Outflow'' Ed. F. Malbet \& A. Castets, p145

\reference{lef96}
Lefloch, B., Eisl\"offel, J.,\& Lazareff, B. 1996,  A\&A, 313, L17

\reference{mat90}
Mathis, J.S. 1990, ARAA, 28, 37

\reference{mrn77}
Mathis, J.S., Rumpl, W., \& Nordsieck, K.H. 1977, ApJ 217, 425

\reference{anc97}
Noriega-Crespo, A. 1997, in ``Herbig-Haro Flows and the Birth of Low Mass 
Stars'' Ed. B. Reipurth \& C. Bertout, p103

\reference{anc97b} Noriega-Crespo, A., Van Buren, D., Cao, Y., Dgani, R. 1997,
AJ, 114, 837

\reference{oh94} Ossenkopf, V., Henning, Th. 1994, A\&A, 291, 943

\reference{smth95} Smith, M.D. 1995, A\&A, 296, 789

\reference{sut97}
Sutter, G., Smith, M.D., Yorke, H.W., \& Zinnecker, H. 1997, A\&A, 318, 595

\reference{tst98} Testi, L., Private Communication 

\reference{wol91}
Wolfire, M.G., \& K\"onigl, A., 1991, ApJ, 383, 205

\end{references}
\end{document}